\documentclass[aps,pra,twocolumn,superscriptaddress,nofootinbib]{revtex4-2}

\usepackage[T1]{fontenc}
\usepackage{amsmath}
\usepackage{amssymb}
\usepackage{graphicx}
\usepackage{xcolor}
\usepackage{hyperref}

\begin{document}

\title{Was Schr{\"o}dinger ever a determinist?}

\author{Flavio Del Santo}
\affiliation{University of Geneva, Group of Applied Physics, Rue de l'\'Ecole-de-M\'edecine 20, Geneva 1211, Switzerland}
\affiliation{Constructor University, Bremen, Germany}

\author{Nicolas Gisin}
\affiliation{University of Geneva, Group of Applied Physics, Rue de l'\'Ecole-de-M\'edecine 20, Geneva 1211, Switzerland}
\affiliation{Constructor University, Bremen, Germany}

\begin{abstract}
\noindent Erwin Schr{\"o}dinger is often portrayed as a reactionary who resisted the indeterminism introduced by quantum mechanics, but on closer inspection his views on determinism prove to be more complex and more radical.
\end{abstract}

\maketitle

A hundred years after the formulation of the fundamental law of quantum mechanics, Schr{\"o}dinger's equation \cite{Schrodinger1926}, quantum physics still rests on a fundamental tension: indeterministic measurement outcomes coexist with the smooth, time-reversal-invariant, deterministic evolution of Schr{\"o}dinger's wave function. While Heisenberg's uncertainty principle and, later, Bell inequality violations led physicists to adopt a fundamentally probabilistic description of nature, Schr{\"o}dinger's equation appeared to reintroduce, to a certain extent, the classical categories of continuity and deterministic causality.

For proponents of the predominant Copenhagen interpretation, in fact, Schr{\"o}dinger---alongside Albert Einstein, Louis de Broglie, and later David Bohm---became associated with a certain ``conservative'' resistance to the revolutionary implications of the new quantum theory. Yet portraying Schr{\"o}dinger as a committed determinist is historically misleading (as well as Einstein \cite{Stone2016} and Bohm \cite{DelSantoKrizek2025}, for that matter). As a student shaped by the so-called ``Vienna indeterminism'' of Ludwig Boltzmann and, especially, his mentor Franz S. Exner \cite{Stoltzner1999}, Schr{\"o}dinger was exposed early on to the idea that ``everything that happens in nature is the result of random events'' \cite{Exner1909}.

It is little known that, in the following years, Schr{\"o}dinger embraced a fundamentally indeterministic view, arguing that physics had always been indeterministic before and independently of quantum theory \cite{Hanle1979}. Since his appointment as professor at the University of Zurich in 1922, Schr{\"o}dinger made his position on indeterminism public, acknowledging and further developing the view inherited from Exner that ``Nature's laws are of thoroughly statistical character'' and that ``the assertion of determinism was certainly possible, yet by no means necessary, and when more closely examined not at all very probable'' \cite{Schrodinger1929}.

Yet, it is in his essay ``Indeterminism in Physics'' \cite{Schrodinger1931} that Schr{\"o}dinger most explicitly sets out his commitment to indeterminism, lamenting not that Heisenberg's views were untenable, but rather that they came too late, since the indeterministic character of physics could already have been recognized long before the advent of quantum theory: ``Why did nobody say, forty or fifty years ago, that modern physics (modern as it was then), was compelled to give up causality and determinism?'' \cite{Schrodinger1931}. This view of classical indeterminism was later independently rediscovered in 1955 by another founding father of quantum physics associated with the school of Copenhagen, Max Born, the proponent of the probabilistic interpretation of the wave function \cite{Born1969}.

In his essay, Schr{\"o}dinger advances three arguments in favor of indeterminism in physics, each of which may appear strikingly at odds with the philosophical views commonly associated with the equation that bears his name:

\textit{(i)} The first argument challenges the mathematical idealization of determinacy in the initial conditions. Schr{\"o}dinger argues that identical initial conditions do not lead to identical outcomes, but only to the same statistics, both in classical and in quantum physics: ``classical mechanics itself is indeterminate. The opposite is generally asserted; but this is due merely to an artifice.'' \cite{Schrodinger1931} The ``artifice'' he refers to is the assumption of infinitely precise velocities as part of the initial conditions in classical equations of motion: ``possibly, however, this is incorrect''---he continues---``possibly this mathematical process of approach to the limit [of the incremental ratio $\Delta x/\Delta t$], which was specially invented by Newton for mechanical purposes, is inadmissible'' \cite{Schrodinger1931}.

\textit{(ii)} The second argument for indeterminism concerns discontinuity and irreversibility. Schr{\"o}dinger suggests that the laws of statistical mechanics may not be merely approximate expressions emerging from underlying deterministic laws, but fundamental in their own right, criticizing determinism as the ``fundamental dogma of classical physics'', an ``ancient custom'', a ``philosophic prejudice'' or an ``a priori belief''. On the contrary, he states that trajectories may be characterized not by exact laws but ``by an appropriate game of chance'': ``As soon as the great majority or possibly all of these laws are seen to be of statistical nature, they cease to provide a rational argument for a retention of determinism'', he claims \cite{Schrodinger1931}. Schr{\"o}dinger then relates this statistical argument to irreversibility, affirming that ``nature is, for the most part, irreversible and one-directional''.

\textit{(iii)} In his final argument, Schr{\"o}dinger links quantum indeterminacy to the abandonment of the concept of trajectory and to the essential role of discontinuity. Remarkably, he can be seen here as anticipating the notion of ``information'' in physics (he does not use the term ``information'', which was formalized only in 1948 by C. Shannon), suggesting that nature may be understood as a discrete sequence of yes--no answers, representable as strings of 0s and 1s. He argues that the information content of nature must, in principle, remain finite, whereas its complexity would require an infinite amount thereof to specify all variables. On this basis, he concludes that indeterminism appears as the more compelling assumption.

The decline of the deterministic dogma is today more widely accepted, certainly thanks to the century-long success of quantum theory. This has led several physicists to reconsider whether physics was ever deterministic, even at the classical level, also in light of the finiteness of information \cite{Born1969,Dowek2013,Drossel2015,DelSantoGisin2019,DelSantoGisin2025}. This development brings quantum and classical physics conceptually closer together, in a way that Schr{\"o}dinger may have already foreseen.

History is written by the winners, as the proverb goes. If one follows this somewhat belligerent narrative of scientific development, Schr{\"o}dinger---who was indeed always part of the opposition to the emerging orthodoxy---has often been portrayed as reluctant to accept the revolutionary implications of quantum physics, including an alleged discomfort with the end of causal determinism. Today, realists about the wave function, such as Everettians, emphasize the fundamental status of the Schr{\"o}dinger equation, thereby treating reversibility and determinism as ineluctable features. Yet, on the contrary, Schr{\"o}dinger may have been so radical in his indeterminism that the significance of his thought may have eluded recognition for a century.

\end{document}